\documentclass[11pt,twoside]{article}
\usepackage{CAGN2019}
\usepackage{graphicx}

\usepackage[T1]{fontenc} % Computer Modern (CM) fonts

\usepackage{latexsym}
\usepackage{verbatim}

\usepackage{ifpdf}  
\ifpdf  
      \DeclareGraphicsExtensions{.pdf,.png,.jpg}  
\else  
      \DeclareGraphicsExtensions{.eps}  
\fi 

\usepackage{comment}

\usepackage{graphicx}
\graphicspath{{figs/}}
\DeclareGraphicsExtensions{.pdf, .eps}
\usepackage[multidot]{grffile}

\usepackage{xspace}
\usepackage[usenames, dvipsnames]{xcolor}

\usepackage{alltt}
\usepackage{ulem} 

\newcommand{\hii}{H\thinspace\textsc{ii}\xspace}

\newcommand{\Ha}{\ifmmode \text{H}\alpha \else H$\alpha$\fi\xspace}
\newcommand{\Hb}{\ifmmode \text{H}\beta \else H$\beta$\fi\xspace}
\newcommand{\Hd}{\ifmmode \text{H}\delta \else H$\delta$\fi\xspace}
\newcommand{\neiii}{\ifmmode [\text{Ne}\,\textsc{iii}] \else [Ne~{\scshape iii}]\fi\xspace}
\newcommand{\Neiii}{\ifmmode [\text{Ne}\,\textsc{iii}]\lambda 3869 \else [Ne~{\scshape iii}]$\lambda 3869$\fi\xspace}
\newcommand{\oii}{\ifmmode [\text{O}\,\textsc{ii}] \else [O~{\scshape ii}]\fi\xspace}
\newcommand{\Oii}{\ifmmode [\text{O}\,\textsc{ii}]\lambda 3726 + \lambda 3729 \else [O~{\scshape ii}]$\lambda 3726 + \lambda 3729$\fi\xspace}
\newcommand{\Oiiit}{\ifmmode [\text{O}\,\textsc{iii}]\lambda 4363 \else [O~{\scshape iii}]$\lambda 4363$\fi\xspace}
\newcommand{\heii}{\ifmmode \text{He}\,\textsc{ii} \else He~{\scshape ii}\fi\xspace}
\newcommand{\Heii}{\ifmmode \text{He}\,\textsc{ii}\lambda 4686 \else He~{\scshape ii}$\lambda 4686$\fi\xspace}
\newcommand{\ariv}{\ifmmode [\text{Ar}\,\textsc{iv}] \else [Ar~{\scshape iv}]\fi\xspace}
\newcommand{\Ariv}{\ifmmode [\text{Ar}\,\textsc{iv}]\lambda 4711 + \lambda 4740 \else [Ar~{\scshape iv}]$\lambda 4711 + \lambda 4740$\fi\xspace}
\newcommand{\Niit}{\ifmmode [\text{N}\,\textsc{ii}]\lambda 5755 \else [N~{\scshape ii}]$\lambda 5755$\fi\xspace}
\newcommand{\hei}{\ifmmode \text{He}\,\textsc{i} \else He~{\scshape i}\fi\xspace}
\newcommand{\Hei}{\ifmmode \text{He}\,\textsc{i}\lambda 5876 \else He~{\scshape i}$\lambda 5876$\fi\xspace}
\newcommand{\nii}{\ifmmode [\text{N}\,\textsc{ii}] \else [N~{\scshape ii}]\fi\xspace}
\newcommand{\niis}{\ifmmode [\text{N}\,\textsc{ii}]_S \else [N~{\scshape ii}]$_S$\fi\xspace}
\newcommand{\Nii}{\ifmmode [\text{N}\,\textsc{ii}]\lambda 6584 \else [N~{\scshape ii}]$\lambda 6584$\fi\xspace}
\newcommand{\oiii}{\ifmmode [\text{O}\,\textsc{iii}] \else [O~{\scshape iii}]\fi\xspace}
\newcommand{\oiiis}{\ifmmode [\text{O}\,\textsc{iii}]_S \else [O~{\scshape iii}]$_S$\fi\xspace}
\newcommand{\Oiii}{\ifmmode [\text{O}\,\textsc{iii}]\lambda 5007 \else [O~{\scshape iii}]$\lambda 5007$\fi\xspace}
\newcommand{\sii}{\ifmmode [\text{S}\,\textsc{ii}] \else [S~{\scshape ii}]\fi\xspace}
\newcommand{\Sii}{\ifmmode [\text{S}\,\textsc{ii}]\lambda 6716 + \lambda 6731 \else [S~{\scshape ii}]$\lambda 6716 + \lambda 6731$\fi\xspace}
\newcommand{\ariii}{\ifmmode [\text{Ar}\,\textsc{iii}] \else [Ar~{\scshape iii}]\fi\xspace}
\newcommand{\Ariii}{\ifmmode [\text{Ar}\,\textsc{iii}]\lambda 7135 \else [Ar~{\scshape iii}]$\lambda 7135$\fi\xspace}
\newcommand{\siii}{\ifmmode [\text{S}\,\textsc{iii}] \else [S~{\scshape iii}]\fi\xspace}
\newcommand{\Siii}{\ifmmode [\text{S}\,\textsc{iii}]\lambda 9069 \else [S~{\scshape iii}]$\lambda 9069$\fi\xspace}
\newcommand{\Siiit}{\ifmmode [\text{S}\,\textsc{iii}]\lambda 6312 \else [S~{\scshape iii}]$\lambda 6312$\fi\xspace}
\newcommand{\rOii}{ \ifmmode [\text{O}\,\textsc{ii} ]\lambda 3726/3729 \else [O~{\scshape  ii}]$\lambda 3726/3729$\fi\xspace}
\newcommand{\rOiii}{\ifmmode [\text{O}\,\textsc{iii}]\lambda 4363/5007 \else [O~{\scshape iii}]$\lambda 4363/5007$\fi\xspace}
\newcommand{\rAriv}{\ifmmode [\text{Ar}\,\textsc{iv}]\lambda 4740/4711 \else [Ar~{\scshape iv}]$\lambda 4740/4711$\fi\xspace}
\newcommand{\rNii}{ \ifmmode [\text{N}\,\textsc{ii} ]\lambda 5755/6584 \else [N~{\scshape  ii}]$\lambda 5755/6584$\fi\xspace}
\newcommand{\rSiii}{\ifmmode [\text{S}\,\textsc{iii}]\lambda 6312/9532 \else [S~{\scshape iii}]$\lambda 6312/9532$\fi\xspace}
\newcommand{\rSii}{ \ifmmode [\text{S}\,\textsc{ii} ]\lambda 6731/6717 \else [S~{\scshape  ii}]$\lambda 6731/6717$\fi\xspace}

\newcommand{\Opp}{O$^{++}$}

\begin{document}
% please do not un-comment the next line
% \input{../../proceeding-book/expages.tex}\setpagenumber{1}

\vskip 1.0cm
\markboth{G. Stasi\'nska}{Strong line abundances}
\pagestyle{myheadings}
%
%%%%  USE THE LINE THAT DESCRIBES THE CHARACTER OF YOUR WORK %%%%%%
%
\vspace*{0.5cm}
\parindent 0pt{Invited Review}
%\parindent 0pt{Poster}

%\vskip 0.3cm

\vspace*{0.5cm}
\title{Can we believe the strong-line abundances 
in giant \hii regions and emission-line galaxies?
}

\author{Gra\.zyna Stasi\'nska $^1$}
\affil{$^1$LUTH, CNRS, Observatoire de Paris, PSL University, France\\
}

\begin{abstract}
This review is not a compendium of strong-line methods to derive abundances in giant \hii regions. It is mostly intended for readers who wish to use such methods  but do not have a solid background on the physics of \hii regions. It is also meant to encourage those using abundance results published in the literature to think more thoroughly about the validity of these results.


\end{abstract}

\section{Introduction}

The knowledge of the chemical composition of the interstellar medium of galaxies is fundamental for our understanding of their chemical evolution. The royal way to derive the chemical abundances of the elements is considered to be the use of the so-called direct method, which is based on the use collisionally  excited emission lines to compute the gaseous elemental abundances  with respect to hydrogen. It involves the measurement of the physical conditions (temperature and electron density distribution) in the emitting plasma. This method requires accurate measurements of weak temperature-sensitive lines, reliable atomic data involved in the procedure, reliable ionisation correction factors for unobserved ions and an estimate of the element depletion in dust grains. The major problem lies in the fact that abundances derived from optical recombination lines in \hii regions are larger than those derived from collisionally excited lines typically by a factor of two. This decades-old problem has so far no generally accepted solution. Until it is solved, element abundances in \hii regions cannot be regarded as completely secure, even if important progress has been achieved at all steps of the procedure to derive them.

The so-called strong line methods to determine abundances in giant \hii regions appeared in 1979. \cite{pageletal79} and \cite{alloinetal79} were the first to propose quantitative estimates of the oxygen abundance in absence of detailed plasma diagnostics, by extrapolating observed trends between strong line ratios and oxygen abundance using the few relevant photoionisation models available at that time. Since then, with the help of extent photoionisation model grids,  numerous high quality observations confirmed that giant \hii regions form basically a one-parameter sequence driven by their `metallicity'\footnote{In the following, the term `metallicity' will be used to designate the oxygen abundance, as is ubiquitously the case  in the literature on the subject.}. 

One could have thought that, with the continuously improving quality of \hii region spectra, strong-line methods would become obsolete. The reverse is happening, due to the deluge of spectroscopic data for \hii regions in close-by galaxies as well as in high-redshift galaxies. Most of the time, these data either are of insufficient spectral resolution and signal-to-noise to provide a reliable determination of electron temperatures (especially at high metallicities)  or they cover a very limited wavelength range allowing the measurement of only a few selected strong lines. 

The purpose of this presentation is not to review the many strong-line methods that are nowadays available as these have been  amply presented in the literature \citep[see e.g.][] {stasinskaetal12, biancoetal16, maiolinomanucci19}, but rather to draw attention to problems worth remembering. Due to space limitations, references in this text are kept to a minimum, and I apologise to all those who contributed to the subject and do not have their work quoted here. More complete sets of references are given in the papers mentioned above.

\section{Strong line methods: the idea}
\label{principles}

Since, in giant \hii regions, strong-line intensity ratios have been shown to present some trends with abundances derived from temperature-based methods, the idea came up that strong lines alone would be sufficient to derive the metallicities\footnote{We neglect the issue of reddening which can easily be accounted for if both \Ha and \Hb are measured.}. This is true -- only statistically, of course -- provided some caution is exercised. The first difficulty is that \textit{any} ratio of the kind $I_C/I_R$ where $I_C$ is the intensity of a collisionally excited line and $I_R$ that a hydrogen recombination line is double-valued with respect to metallicity. One thus has to choose which of the two regimes -- high or low metallicity -- is appropriate for the object under study. The choice is either based on external arguments (e.g. \hii regions in the central parts of galaxies are supposed to be metal-rich,  low-mass star-forming galaxies are supposed to be metal-poor, etc.) or on an argument based more or less directly on the N/O ratio (expected to be large at high metallicities). The second difficulty is that \hii regions do not strictly form  a one-parameter sequence. While it has been shown that there is a relation between the metallicity, the hardness of the ionising radiation field and the mean ionisation parameter of the \hii regions -- which are the main parameters determining their spectrum -- the relation likely suffers some dispersion. For example, giant \hii regions of same metallicity can be ionised by star clusters of different ages. In addition, other parameters (nebular geometry, dust content, ionising photon leakage, etc.) also play a role, even if not dominant. Their role should be evaluated, especially when one is looking for  metallicity differences between two classes of similar objects.  For example \hii regions in  spiral arms versus \hii regions between  spiral arms.

\section{The importance of calibration}
\label{calibration}

Any strong-line method needs to be calibrated, either using data sets with  abundances from direct methods, or using photoionisation models, or a combination of both.

If based on observations, the calibrating sample must represent well the \textit{entire} family of \hii regions to be studied. For example, calibrations based on local, high-density \hii regions should no be used for giant \hii regions. Calibrations based on giant \hii regions should not be used to derive the global metallicity of large portions of galaxies, such as observed in the Sloan Digital Sky Survey. 

One of the drawbacks of observational calibrations  is that the weak emission lines  needed to determine the electron temperature cannot be observed at very high metallicities because of their strong dependence on the excitation energy\footnote{Even when \Oiiit can be measured but is very weak, it may lead to a completely wrong temperature diagnostic for the entire \Opp zone, as shown by \cite{stasinska05}.}. In addition, it is likely that the calibration samples are biased towards the highest temperatures at a given metallicity. The recent use of stacked spectra of emission-line galaxies allows one to obtain high signal-to-noise ratios even for very weak lines \citep{liangetal07, andrewsmartini13, curtietal17, bianetal18} and opens the way to observational calibrations for high metallicities but the results may depend on the stacking parameters, which have to be carefully chosen, as shown by \cite{brownetal16}.

Using model grids, one is free from the selection biases mentioned above since photoionisation models can be produced at any metallicity. But photoionisation models are a simplified representation of reality and require ingredients that are not fully known. For example, the spectral energy distribution of the ionising radiation relies on stellar atmosphere models which are difficult to validate observationally, as well as on a correct description of the stellar initial mass function and stellar evolution, for which rotation and the effect of binarity have been introduced only recently. The geometrical distribution of the nebular gas does play a role which is generally ignored. The ratios of elemental abundances with respect to oxygen are not constant: some dispersion is expected, especially for nitrogen and carbon, since their production sites are not identical. The presence of dust grains affects both the chemical composition of the gas phase and the line emission -- either directly (because of depletion) or indirectly (because of the effect of grains on the ionization structure and on the nebular temperature). Calibrations based on different grids of models may lead to different results.

Finally, the role of atomic data cannot be ignored. They play a crucial role in direct abundance determinations as well as in photoionisation models. Even recently, some important atomic data have suffered changes (e.g. the collision strengths for \oiii by \cite{palayetal12}, later revised by \cite{storeyetal14} and \cite{tayalzatsarinny17}.

\section{Various approaches}
\label{approaches}

\subsection{Indices}
\label{indexes}

The easiest way to obtain metallicities from strong lines is to use analytic expressions based on some indices, such as the famous (\oiii + \oii)/\Hb \citep{pageletal79} or \oiii/\nii \citep{alloinetal79} (which have been recalibrated many times). This method is still being used because of its simplicity and minimal observational requirements. Other indices have been also proposed since then. However, it must be kept in mind that methods based on indices are especially prone to biases which may lead to erroneous astronomical inferences.

This method has been extended to the use of two indices to account for a second parameter (namely the ionisation level) that influences the \hii regions spectra \citep{mcgaugh94, pilyugin01}. It is customary to compute the error bars taking into account \textit{only }the uncertainties in the line intensities and not the intrinsic statistical uncertainty of the method.

\subsection{Interpolation}
\label{interpolation}

A more elaborate way to obtain abundances from strong lines is to use an iterative approach to interpolate in a grid of models \cite[e.g.][]{kewleydopita02}. In this very paper, the uncertainties are evaluated by comparison of results using several methods.

An even more sophisticated approach is to interpolate within a grid by  $\chi^2$ minimization or a related  technique using selected line ratios \cite[e.g.][]{charlotetal02,perezmontero14}. 

Instead of a grid of models, one can also use real objects for which direct abundances are available, based on the assumption that `\hii regions with similar intensities of strong emission lines have similar physical properties and abundances' \citep{pilyuginetal12} -- which is not entirely true.

\subsection{Bayesian methods}
\label{bayesian}

A common belief is that, when many lines are used (i.e. not only the few classical strong lines such as \oiii, \oii, \nii with respect to \Ha or \Hb) the resulting abundances will be more accurate. This is not necessarily correct. Considering additional lines from other elements may give more information on the physical conditions of the nebulae, but at the same time introduce a dependence on abundance ratios that are not varied in the reference grid of models.  In addition, lines such as \sii, \neiii, \ariii, \siii being often relatively weak may add noise or biases to the procedure. Nevertheless, a treatment of the available information in a Bayesian framework may lead to reliable inferences \cite{chevallardcharlot16}.
Bayesian methods are becoming more and more popular to determine chemical abundances \citep{blancetal15, valeasarietal16, thomasetal18}. However it is important to realise that each author uses his own philosophy and the results -- including the characterisation of the uncertainties -- depend on the priors.

\subsection{Machine learning}
\label{machine}

The next step, which has already been taken, is to use machine learning techniques to derive abundances -- together with other properties \citep{uccietal17, uccietal18, uccietal19, ho19}. Such techniques try to make the best use of the information encoded in the observational data by considering not only some pre-selected emission lines, but the entire observation set -- without necessarily understanding the underlying physics. Machine learning techniques are widely used in many areas of science -- including human sciences -- when analysing huge amounts of data which depend on a large number of parameters whose role is not fully understood and which are sometimes not even identified. It is not clear what is their advantage in the case of abundance determinations  except, perhaps, their ability to treat a colossal amount of data in a short  time. Regarding nebulae and their chemical composition, we know, in principle, what governs the production of the emission lines and even what are the steps undergone by the photons generated in the nebulae before being recorded by our observational devices. 
In fact, machine learning approaches which ignore the physics of the studied phenomena may lead to very strange inferences. For example in the on-line accompanying data of the study by \cite{uccietal18}, it is stated that `the 6 most informative features are lines connected to H and He transitions (i.e. \Hb, \Hd, \Hei) and 3 metals lines (i.e. \Oii, \Neiii, and \Ariii). [...] the two most informative lines, \Hd and \Ariii, sum up to over 40\% of the total feature importance'. Such a statement will of course puzzle anyone familiar with the physics of nebulae!

\section{Biases, misinterpretations and mistakes}
\label{biases}

The fact that strong-line methods have become more and more sophisticated over the years should not prevent us from keeping a critical eye on their results and on their astrophysical inferences:  a blind confidence is not recommended.

\subsection{Very indirect abundance estimates}
\label{indirect}
The real relation between N/O and O/H is the interstellar medium of galaxies is in itself important to know for understanding the production conditions of nitrogen and the chemical history of galaxies \cite{mollagavilan10}.
 But it is also critical for a proper determination of the oxygen abundance in most strong-line methods. For example, the \nii/\oii index promoted by \cite{kewleydopita02} to estimate O/H is actually directly related to N/O. The same is true for the \oiii/\nii index, although the latter is also linked to the excitation state of the nebula  which, empirically, is linked to metallicity. Even in the case when the metallicity derivation is based on a library of models, the assumed N/O versus O/H relation plays an important role. This has already been mentioned (Section \ref{huge}) for the oxygen abundances based on the grid of \cite{charlotlonghetti01} but is also true for many other works \citep[see e.g.]{dopitaetal16}. So far,   N and O are considered independently only in the grids of \cite{perezmontero14} and \cite{valeasarietal16}. 

\subsection{Models are not perfect}
\label{perfect}

Photoionisation models composing the grids are based on very simple geometries for the gas and the dust. Their predicted  \Oiiit/\Oiii line ratios are often smaller than observed (this explains why strong-line methods often give larger abundances that temperature-based methods). This may be due to a  density distribution that is too schematic, but also to an input assumed  spectral energy distribution  (SED) of the ionising photons that is too soft, or to the presence of temperature inhomogeneities of unknown nature. Note that the long-lasting problem of the presence of \heii lines in metal-poor \hii regions seems to be solved by the presence of high-mass X-ray binaries (not yet routinely considered in most models for giant \hii regions) although contributions from shocks are not excluded \citep{schaereretal19}.  

Another problem is that, so far, strong-line methods based on model grids  consider only ionisation-bounded cases while photon leakage from star-forming complexes is now well-documented. This issue should be considered for giant \hii regions although, in the case of integrated observations of emission-line galaxies, one also need to consider the diffuse ionised medium (see Sect. \ref{entire}).

\subsection{Gigantic grids are not sufficient}
\label{huge}

In methods based on large model grids, one may think that the grids are so extensive that they encompass all the possible situations. This is far from true. 
If the grid does not include a reasonable representation of \textit{all} that is seen in nature, the abundance results may be biased. 

For example, the oxygen abundances derived by \cite{tremontietal04} based on a library of $2 \times10^5$ models by \cite{charlotlonghetti01} were shown by \cite{liangetal07} to be strongly overestimated at the high metallicity end. Their library assumed a unique value of N/O for all metallicities and the fitting was made with a $\chi^2$ technique summing on all the available strong lines, so that a mismatch of the \nii/\Ha ratio between observations and models required a modification of the O/H ratio.

Abundances derived from model grids can be incorrect just because the method is inadequate. For example, Fig. C4 in \cite{valeasarietal16} demonstrates that the {\scshape hii-chi-mistry} code by \cite{perezmontero14} may give erroneous oxygen abundances when excluding \Oiiit from the fitting procedure.

\subsection{Fake relations}
\label{fake}

Empirical expressions to derive abundances from line ratios can artificially reduce an intrinsic scatter. This is exemplified in Fig. C5 of \cite{valeasarietal16} in the case of the ON calibration from \cite{pilyuginetal12}. When applied to a sample models where N/O and O/H are loosely correlated, the ON calibration tightens the resulting relation significantly.

Using the {\scshape hii-chi-mistry} code for spaxel-analysis of integral-field spectroscopic observations of the blue compact dwarf galaxy NGC 4670 \cite{kumarietal18} find that N/O anti-correlates with O/H. This result is in clear contrast with previous studies on the behaviour of N/O versus O/H, and likely stems from the fact that the observational uncertainties are very large leading to an apparent anti-correlation between N/O and O/H.

As a matter of fact, some of the  methods to obtain the metallicity are \textit{only indirectly} linked to this parameter. They work just because, in astrophysical situations, other parameters are related to metallicity. This is the case of the \nii/\oii ratio mentioned above,  or of \nii/\Ha \citep{denicoloetal02}. The latter is very convenient because it requires only a very narrow spectral range and does not depend on reddening. But its is essentially  driven by N/O and the nebular excitation -- which both happen to be astrophysically related to the metallicity. 

\subsection{Degeneracies}
\label{degeneracies}

Because the ratio of collisional and recombination  lines is degenerate 
with respect to metallicity, a way must be found to distinguish high metallicity from low metallicity objects. As mentioned in Section~\ref{principles}, this is generally done using arguments of astrophysical nature. The fact that N and O follow a different chemical evolution in galaxies is obviously of great help. However, there may be outliers, and those can be of particular astrophysical interest. In their bayesian approach {\scshape BOND}, \cite{valeasarietal16} devised a way to 
find whether an object is oxygen-rich or oxygen-poor, independently of N/O. It relies on the semi-strong lines \Neiii and \Ariii whose ratio   depends strongly on the electron temperature and weakly on ionisation conditions and abundance ratio. Indeed,  the \Ariii and \Neiii lines have different excitation thresholds (1.7 and 3.2 eV, respectively), therefore different dependencies on the electron temperature;  argon and neon are two primary elements and  both are rare gases not suspected of dust depletion, so their abundance ratio in the gas phase is expected to be constant. The \Ariii and \Neiii  lines do not arise exactly from the same zone, the ionisation potentials of the parent ions being different, but the \oiii/\oii  ratio helps figuring out what the ionisation is. {\scshape BOND} allows for a possible small deviation of Ne/Ar from the standard cosmic value by applying appropriate weights to their lines.

It must be noted, however, that {\scshape BOND} will probably not work correctly in the case of large observational errors in the strong line fluxes (say 20 percent).

\subsection{The problem when comparing two samples and the role of the hardness of the SED}
\label{twosamples}

Some claims about the metallicity and its connexion with other galaxy properties could be wrong by actually arising from the fact that different physical conditions in the galaxies -- not accounted for in the method used to obtain the metallicity -- mimic a change in abundance. Such could be the case  for the surprising finding by \cite{sanchezalmeidaetal09} that oxygen abundances are larger in galaxies with older starbursts.  As showed by \cite{stasinska10} this could be due to a wrong interpretation of the \nii/\Ha index in terms of just oxygen abundance while it is primarily dependent on the hardness of the ionising radiation field which softens with age due to the gradual disappearance of the most massive stars. Similarly, the claim that luminous infrared galaxies (LIRGs)  have smaller metallicities than local star-forming galaxies \citep{rupkeetal0} should be reexamined having in mind that the hardness of the ionising radiation is likely different in the two categories of galaxies. The same can be said about the claim that metallicity gradients flatten in colliding systems  \citep[see e.g.][]{kewleyetal10}. So far, only the {\scshape BOND} method of \cite{valeasarietal16}   determines the oxygen abundance trying to account for the characteristics of the ionising radiation field. The semi-strong line \Hei plays a key role in the process.

\subsection{Do we get what we want?}
\label{want}

In a recent study using neural networks to obtain galaxy metallicity from three-colour images, \cite{wuboada19} suggest that their approach has `learned a representation of the gas-phase metallicity, from the optical imaging, beyond what is accessible with oxygen spectral lines'. This \textit{a priori} surprising statement is based on the fact that with their method they find a  mass-metallicity relation for galaxies that is tighter than the one derived by \cite{tremontietal04}. They further comment that possibly their method `is measuring some version of metallicity that is more fundamentally linked to the stellar mass, rather than [the metallicity] derived from oxygen spectral lines'. Indeed, galaxy colours are not only determined by the metallicities of their stellar populations (which are different from the metallicities of their interstellar medium) but also by their ages, and both are known to be correlated with stellar mass. This interesting case should incite us to remain cautious about the interpretation of the oxygen abundance determined by any method: is it really the present-day oxygen abundance that is being measured or something else related to it?

\section{Metallicities of entire galaxies and the importance of the DIG}
\label{entire}

While the methods to derive the chemical composition of the ionised interstellar medium were developed 
for giant \hii regions, they are often applied to entire galaxies or, at least, to a large fraction of their volume \citep[e.g][]{tremontietal04, mannuccietal10}. This is not only because observations are now available for galaxies at high redshifts. The `characteristic' metallicity of a galaxy is an important constraint for chemical evolution models. Past studies have tried to determine it in nearby galaxies from the study of their \hii regions \citep{zaritskyetal94, moustakaskennicutt06}.

Integrated spectra of galaxies combine the spectra of hundreds of \hii regions of various sizes, ages and metallicities. They also contain emission from the diffuse ionized gas (DIG), whose characteristics are different from those of \textit{bona fide} \hii regions \citep{lacerdaetal18}. Integral field spectroscopy of nearby galaxies (etc CALIFA, MANGA, SAMI, MUSE) now allow one to investigate these issues and try to find ways  to account for them for a proper characterisation of the metallicity of nearby and distant galaxies.

The existence of a diffuse component in the ionised medium of galaxies has been acknowledged several decades ago \citep{reynolds90, hoopesetal99}. On the basis of a sample of about 100 H\thinspace\textsc{i}\xspace selected galaxies it has been claimed that diffuse \Ha emission  constitutes roughly 60 per cent of the total \Ha emission, irrespective of the Hubble type \citep{oeyetal07}. In a more detailed study of 391 galaxies from the CALIFA survey \citep{sanchezetal12}, \cite{lacerdaetal18} have shown that the DIG is quite complex. Regions with the lowest equivalent widths of \Ha (called hDIG) are likely ionised by hot low-mass evolved stars (HOLMES) while diffuse zones with  \Ha equivalent widths between 3 and 14 \AA\ (dubbed mDIG) are rather ionised by radiation leaking out from  star-forming complexes (SFc).  The hDIG, mDIG, and SFc contributions to the total \Ha luminosity vary systematically along the Hubble sequence, ranging from about (100, 0, 0) per cent in ellipticals and S0's to (9, 60, 31) per cent in  Sa-Sb's and (0, 13, 87) per cent in later types.

Based on resolved spectroscopic observations of nearby galaxies, a few studies are starting to investigate the effect of the DIG on the abundance determination from emission lines in galaxies \citep{sandersetal17, kumarietal19, erroz-ferreretal19}.

\section{Conclusion}
\label{conclusion}

While strong-line methods are routinely used to estimate the metallicities of giant \hii regions as well as the characteristic of metallicities of local and high-redshift emission-line galaxies, one can easily be fooled by their apparent simplicity. 
Even the latest more sophisticated approaches involve a certain degree of approximation.
The purpose of this text was to draw attention to some pitfalls and to encourage astronomers to consider their results  with appropriate caution.

%Of course, for a meaningful comparison of abundances in various objects, the abundances must be derived using the same indexes and the same calibrations.
%
%no priors vs extreme priors
%
%indexes still useful

%\begin{figure}  %%%%%%%%%%%%%%%%%%%%%%FIGURE 1 %%%%%%%%%%%%%%%%%%%%%%%
%\begin{center}
%\hspace{0.25cm}
%\includegraphics[height=5.0cm]{fig1}
%\caption{Cooking an {\em asado}. Notice the lattice-shaped {\em parrilla}, and
%the proper order of the meat and {\em chorizos}.}
%\label{parrilla}
%\end{center}
%\end{figure}

\acknowledgments 
I thank the organisers for their kind invitation  and deeply acknowledge financial support from FAPESP (processo 11/51680-6) which allowed me to participate in the workshop.

\bibliographystyle{aaabib}
\bibliography{stasinska}

%\references
%%
%\bibitem[Alloin et al.(1979)]{1979A&A....78..200A} Alloin, D., Collin-Souffrin, S., Joly, M., et al.\ 1979, \aap, 78, 200.
%
%Andrews \& Martini, 2013
%
%Bian et al., 2018)
%
%Bianco et al 2016, 
%
%Blanc et al. 2015
%
%Brown et al. (2016).
%
%Charlot \& Longhetti (2000) 
%
%Charlot et al 2002
%
%Chevallard et al 2016
%
% Couto
% 
%Curti et al., 2017
% 
%Denicolo et al 2002
% 
%Dopita et al 2016
%
%Erroz-Ferrer et al 2019,
%
%Ho 2019
%
%Hoopes et al 1999)
%
%Kewley \& Dopita 2002)
%
%Kewley et al 2010
%
%Kumari et al 2018
%
%Kumari et al 2019, 
%
%Lacerda
%
%Liang et al 2007
%
%Maiolino \& Manucci 2018
%
%Mannucci et al 2009
%
%Mc Gaugh 1994, 
%
%Molla
%
%Moustakas \& Kennicutt 2006).
%
%Oey et al. 2007
%
%pagel et al 79
%
%Palay et al. 2012 
%
%Perez-Montero 2014)
%
%Pilyugin 2001)
%
%Pilyugin et al. (2010)
%
%Pilyugin et al. 2012).
%
%Reynolds 1990, 
%
%Rupke et al 2008) 
%
%S\'anchez et al. 2012)
%
%Sanders et al 2017, 
%
%Sanchez-Almeida et al (2009)
%
%Schaerer et al 2019
%
%Stasi\'nska (2006)
%
%Stasi\'nska (2010)
%
%Stasi\'nska et al. 2012, 
%
%Storey et al. 2014 
%
%Tayal \& Zatsarinny 2017).
%
%Thomas et al 2018).
%
%Tremonti et al. (2004) 
%
%Ucci et al 2017
%
%Ucci et al 2018
%
%Ucci et al 2019
%
%Vale Asari e al 2016
%
%Wu \& Boada (2019)
%
%Zaritzky et al 1994
%
%

\end{document}